\documentclass[sigconf]{acmart}

\usepackage{multirow}
\usepackage{cleveref}
\usepackage[table]{xcolor}
\usepackage{subcaption}
\usepackage[table]{xcolor}
\usepackage{fontawesome5}

\definecolor{OIOrange}{HTML}{E69F00} 
\definecolor{OISkyBlue}{HTML}{56B4E9} 

\colorlet{SchlattRow}{OIOrange!18}
\colorlet{OurRow}{OISkyBlue!18}

\newcommand{\CLS}{\texttt{[CLS]}}
\newcommand{\githubrepo}[1]{%
    \begin{center}
        \href{#1}{\small \faGithub \ \texttt{#1}}
    \end{center}
}

\AtBeginDocument{%
  }

\copyrightyear{2026}
\acmYear{2026}
\setcopyright{cc}
\setcctype{by}
\acmConference[CIKM '26]{Proceedings of the 35th ACM International Conference on Information and Knowledge Management}{November 07--11, 2026}{Rome, Italy}
\acmBooktitle{Proceedings of the 35th ACM International Conference on Information and Knowledge Management (CIKM '26), November 07--11, 2026, Rome, Italy}
\acmDOI{10.1145/3799682.3840914}
\acmISBN{979-8-4007-2539-5/2026/11}

\acmSubmissionID{1796}

\begin{document}

\title{Building Better Encoder-only Cross-Encoders: A Controlled Study of Training Strategies for Neural Re-ranking}


\author{Victor Morand}
\authornote{Both authors contributed equally to this research.}
\orcid{0009-0001-5312-2998}
\affiliation{%
  \institution{Sorbonne Université, CNRS, ISIR}
  \city{Paris}
  \country{France}
}
\email{victor.morand@isir.upmc.fr}

\author{Mathias Vast}
\authornotemark[1]
\orcid{0009-0007-4612-717X}
\affiliation{%
  \institution{Sinequa by ChapsVision}
  \city{Paris}
  \country{France}
}
\affiliation{%
  \institution{Sorbonne Université, CNRS, ISIR}
  \city{Paris}
  \country{France}
}
\email{mathias.vast@isir.upmc.fr}

\author{Basile Van Cooten}
\orcid{0009-0002-0234-917X}
\affiliation{%
  \institution{Sinequa by ChapsVision}
  \city{Paris}
  \country{France}
}
\email{bvancooten@chapsvision.com}

\author{Laure Soulier}
\orcid{0000-0001-9827-7400}
\affiliation{%
  \institution{Sorbonne Université, CNRS, ISIR}
  \city{Paris}
  \country{France}
}
\email{laure.soulier@isir.upmc.fr}

\author{Josiane Mothe}
\orcid{0000-0001-9273-2193}
\affiliation{%
  \institution{Université de Toulouse, CNRS, IRIT, CLLE}
  \city{Toulouse}
  \country{France}
}
\email{josiane.mothe@irit.fr}

\author{Benjamin Piwowarski}
\orcid{0000-0001-6792-3262}
\affiliation{%
  \institution{Sorbonne Université, CNRS, ISIR}
  \city{Paris}
  \country{France}
}
\email{benjamin.piwowarski@cnrs.fr}

\renewcommand{\shortauthors}{Morand et al.}

\begin{abstract}

Cross-encoders fine-tuned from Transformer backbones remain the standard for second-stage re-ranking, and recent knowledge-distillation strategies have closed much of the gap with LLM re-rankers. However, these strategies have not been compared under controlled conditions. In particular, it remains unclear how distillation from LLM rankers compares to distillation from strong cross-encoder teachers, or to purely supervised objectives. It is also unclear how much newer backbones (RoBERTa, ELECTRA, DeBERTaV3, ModernBERT) contribute compared to the original BERT.
We run 162 controlled training runs (9 backbones $\times$ 6
objectives $\times$ 3 seeds), spanning pointwise, pairwise, and listwise losses
with both human labels and two distillation signals, and evaluate on
TREC-DL, MS~MARCO dev, BEIR, LoTTE, and Robust04. We find that objectives emphasizing relative
comparisons—pairwise MarginMSE and listwise InfoNCE—consistently outperform
alternative objectives, including more complex listwise LLM distillation, across all backbones, and switching objective yields
gains comparable to moving up one backbone size tier. A controlled
disentanglement further shows that, once the negative-sampling pool is
matched, even a simple pairwise Hinge loss with ColBERTv2 hard negatives
matches---and on out of domain beats---listwise LLM distillation, indicating that the
quality of the negatives is at least as important as the choice of loss.
We release all 162 trained models  on
HuggingFace\footnote{https://huggingface.co/collections/xpmir/reproducing-cross-encoders} and a unified training codebase.

\githubrepo{https://github.com/xpmir/cross-encoders}
  
\end{abstract}




\maketitle


\section{Introduction}

The two-stage retrieve-then-re-rank pipeline is a standard pattern in Information Retrieval~(IR)~\cite{liuLearningRankInformation2011}. This paradigm was popularized by \citet{monobert}, who demonstrated that high-performance rankers could be obtained by fine-tuning encoder-only backbones, such as BERT \cite{bert}, using cross-entropy on MS~MARCO~\cite{msmarco} (MSM) binary relevance labels. This approach swiftly expanded to include encoder-decoder \cite{monot5, rankT5} and decoder-only transformer architectures \cite{rankLlama}. The rapid scaling of decoder-only LLMs led to promising results with backbones such as Qwen3 \cite{yang2025qwen3technicalreport}, notably when used in listwise re-ranking configurations \cite{wangJinarerankerv3LastNot2025}.


Despite these promising zero-shot results from decoder-only LLMs, recent studies indicate that traditional encoder-only architectures outperform decoder-only architectures when evaluated at a fixed parameter count \cite{ettin, Dejean2024ATC}. This efficiency-to-performance ratio motivates our focus on encoder-only re-rankers for this comparative study, as they remain the most viable option for high-throughput production environments.

Advances on encoder-only re-rankers have followed two complementary directions.
First, stronger training strategies have been proposed, especially via knowledge distillation \cite{hinton2015distillingknowledgeneuralnetwork}, where a strong teacher provides soft supervision signals in place of binary labels. \citet{hofstatterImprovingEfficientNeural2020} introduced distillation from an ensemble of cross-encoder teachers towards student retrievers. More recently, this line of research shifted toward using Large Language Models (LLMs) as teachers, given their promising zero-shot ranking abilities \cite{sunChatGPTGoodSearch2023, twolar,schlattRankDistiLLMClosingEffectiveness2025}. \citet{schlattRankDistiLLMClosingEffectiveness2025} use listwise distillation objectives to achieve further gains, though their experimentation is limited to a single backbone \cite{clark2020electra}.

Second, increasingly strong pre-trained encoder backbones have emerged (e.g., RoBERTa~\cite{liu2019roberta}, ELECTRA~\cite{clark2020electra}, DeBERTaV3~\cite{he2021debertav3}, and ModernBERT~\cite{ettin}), raising the question of whether reported gains come primarily from the objective/teacher or from the underlying architecture. These advances are difficult to compare directly because existing studies \cite{schlattRankDistiLLMClosingEffectiveness2025} vary multiple factors at once: backbone, loss formulation, teacher model, candidate generation, and evaluation benchmarks. This makes it hard to isolate robust design choices. For example, training pipeline choices such as how negatives are mined and which retriever is used can substantially affect re-ranker outcomes \cite{rethink_ce_training,pradeepSqueezingWaterStone2022}. However, the relative performance of these strategies---such as distillation from strong cross-encoder teachers---has not been systematically evaluated.


Motivated by this gap, we provide a controlled, unified methodology for training and evaluating cross-encoder re-rankers using various strategies and backbone architectures, and release code with weights and training metrics for all trained models to support future work.
Accordingly, our benchmark focuses on disentangling the effect of (i) the training objective family, especially the source of the distillation signal, and (ii) the encoder backbone, under a shared evaluation protocol.
Concretely: 

\begin{enumerate}
    \item We compare recent distillation techniques \cite{hofstatterImprovingEfficientNeural2020, schlattRankDistiLLMClosingEffectiveness2025} against supervised losses covering pointwise (BCE~\cite{monobert}), pairwise (Hinge~\cite{macavaney2019ContextualWR}), and listwise (InfoNCE~\cite{caoLearningRankPairwise2007}) objectives trained on raw MSM labels, complementing recent analyses on decoder-only re-rankers \cite{xu2025distillationversuscontrastivelearning}.
    \item We disentangle the contribution of the loss formulation from that of the negative-sampling pool, and show that a simple pairwise Hinge loss paired with ColBERTv2 hard negatives matches---and on OOD beats---listwise LLM distillation, challenging the necessity of expensive distillation pipelines.
    \item We benchmark these objectives on a broad set of encoder-only backbones: BERT~\cite{bert}, RoBERTa~\cite{liu2019roberta}, ELECTRA~\cite{clark2020electra}, DeBERTaV3~\cite{he2021debertav3}, and the Ettin/ModernBERT suite~\cite{ettin}, and find that switching objective yields gains comparable to moving up one backbone size tier.
    \item We evaluate all 162 configurations on a unified suite: MSM dev~\cite{msmarco}, TREC-DL'19/20~\cite{craswell2019overview,craswell2020overview} for in-domain (ID) validation, and BEIR~\cite{thakur_beir_2021}, LoTTE~\cite{santhanam2022colbertv2}, and TREC-Robust04~\cite{Voorhees2004Robust} for out-of-domain (OOD)\footnote{All strategies use the MS~MARCO training data.}.
\end{enumerate}



\section{Related Works}

Prior works on cross-encoder training often vary multiple factors at once---candidate generation, backbone, training protocol, and benchmark suite---making it difficult to isolate the effect of any single design choice \cite{monobert,hofstatterImprovingEfficientNeural2020,schlattRankDistiLLMClosingEffectiveness2025}. Closer to ours, a handful of studies vary one axis carefully while controlling others.

\textbf{Negatives and first-stage retrieval.} \citet{pradeepSqueezingWaterStone2022} study the impact of negative-sampling strategies for cross-encoders while varying the encoder (BERT-base to ELECTRA-base) and the retriever (BM25 vs.\ TCT-ColBERTv2). \citet{Dejean2024ATC} compare LLMs to cross-encoders for re-ranking SPLADE candidates, investigating the choice of first-stage retriever, re-ranker, and pool size. Both fix the loss formulation.

\textbf{Backbone scaling and objectives.} \citet{seetharaman2026scalinglawscrossencoderreranking} study the scaling law of Ettin-based cross-encoders under pointwise, pairwise, and listwise objectives, without considering distillation losses or backbones other than ModernBERT \cite{modernbert}. \citet{xu2025distillationversuscontrastivelearning} compare decoder-only re-rankers trained with InfoNCE \cite{caoLearningRankPairwise2007} (contrastive) or Kullback--Leibler divergence (distillation); their study is limited to Qwen-based re-rankers \cite{yang2024qwen25mathtechnicalreportmathematical}, considers one loss per category, and does not investigate negative sampling.

\textbf{Our positioning.} To the best of our knowledge, this is the first study to jointly vary IR objectives, negative-sampling pools, and encoder backbones under a unified evaluation protocol, offering actionable guidance for practitioners. We separate the discussion of supervision signals and losses (\Cref{section:background}) from the controlled experimental setup (\Cref{sec:framework}), which fixes candidate generation, training protocol, and evaluation benchmarks across all runs.

\section{Background and Notation}\label{section:background}

\textbf{Re-ranking setting and notation.} We consider pointwise cross-encoder re-rankers that assign a relevance score $S_\theta(q,p)$ to a query--passage (or document) pair $(q,p)$ by explicitly modelling their interactions. At inference time, the re-ranker scores a candidate set of $n$ passages $\{p_1, \ldots, p_n\}$ produced by a first-stage retriever, with associated ground-truth labels $y_i \in \{0,1\}$ and predicted logit scores $s_i = S_\theta(q, p_i)$. Re-ranking effectiveness depends on both the scoring model and the candidate set, standardized in \Cref{sec:framework}.

\textbf{Supervised learning-to-rank objectives.}
Cross-encoder training objectives fall into three families \cite{liuLearningRankInformation2011}: \emph{pointwise} objectives treat each $(q,p)$ pair independently (e.g.\ cross-entropy as in \citet{monobert}); \emph{pairwise} objectives optimize relative preferences between a positive and a negative passage for the same query (e.g.\ Hinge loss as used by \citet{macavaney2019ContextualWR}); \emph{listwise} objectives optimize over a full candidate set, e.g.\ an InfoNCE-style formulation following \citet{caoLearningRankPairwise2007}. We formalize the expression of the six objectives used in this work in \Cref{sec:training_strategies}.

\textbf{Distillation as supervision for re-ranking.}
Knowledge distillation \cite{hinton2015distillingknowledgeneuralnetwork} replaces hard relevance labels with signals from a stronger teacher. In re-ranking, supervision can take the form of (i) teacher scores for individual $(q,p)$ pairs, (ii) teacher margins between positive and negative candidates, or (iii) a teacher-induced ranking over a list of candidates. This paper studies (ii) and (iii): the pairwise MarginMSE objective of \citet{hofstatterImprovingEfficientNeural2020}, which distills from an ensemble of cross-encoder teachers, and the listwise DistillRankNet / ADR-MSE objectives of \citet{schlattRankDistiLLMClosingEffectiveness2025}, which distill from an LLM ranker (RankZephyr \cite{pradeep2023rankzephyr}). Their formulations and teacher pipelines are detailed in \Cref{sec:training_strategies}.

\section{Experimental Framework}\label{sec:framework}


To isolate the impact of the training signal / objective from the encoder backbone, we benchmark representative distillation-based strategies from the literature \cite{schlattRankDistiLLMClosingEffectiveness2025,hofstatterImprovingEfficientNeural2020} 
alongside supervised baselines, under a controlled experimental environment. Note that since \citet{hofstatterImprovingEfficientNeural2020} trained dual-encoder students rather than cross-encoders, our use of their released teacher scores and MarginMSE objective for cross-encoder training is a new application.


 

\subsection{Datasets and First-Stage Retriever}

\textbf{Datasets}
Following the literature, we use MS~MARCO~v1 \cite{msmarco} (MSM) to train our models.
Our evaluation framework assesses effectiveness across both in-domain (ID), using the official dev set of MSM, TREC-DL'19 and TREC-DL'20 \cite{craswell2019overview,craswell2020overview}, and out-of-domain (OOD) scenarios, using the 13 publicly available datasets of the BEIR benchmark \cite{thakur_beir_2021} (excluding our training dataset MSM, following previous work \cite{lassance2024spladev3, xu-etal-2025-csplade, wangJinarerankerv3LastNot2025}), the Search version of the 5 LoTTE datasets \cite{santhanam2022colbertv2}, as well as TREC-Robust04 \cite{Voorhees2004Robust}. For each query, we re-rank the top-1000 passages retrieved by SPLADE-v3-DistilBERT \cite{lassance2024spladev3}. This decision is motivated by (1) the prevalence of sparse approaches for first-stage retrieval due to their high efficiency \cite{bm25,formal2021splade,formal2021spladev2,lassance2024spladev3} compared to late-interaction approaches such as ColBERTv2, and (2) the superiority of SPLADE-v3-DistilBERT over BM25 \cite{bm25}. 

\begin{table}[t]
\centering
\caption{Average Recall@1000 results of first-stage retrievers across ID and OOD datasets.}
\label{tab:first-Stage}
\resizebox{0.9 \columnwidth}{!}{
\begin{tabular}{r|c|ccc}
\toprule
\textbf{Retriever} & \textbf{ID} & \textbf{BEIR} & \textbf{LoTTE} &\textbf{OOD} \\
\midrule
BM25 & 77.4 & 75.5 & 80.9 & 76.9 \\
SPLADE-v3-DistilBERT &  \textbf{88.2} & \textbf{79.0} & \textbf{88.2} & \textbf{81.4} \\
\bottomrule
\end{tabular}
}
\end{table}

\noindent
Indeed, preliminary experiments show that SPLADE-v3-DistilBERT significantly outperforms BM25 in terms of Recall@1000 across all benchmarks, achieving an average OOD recall of 81.4, compared to 76.9 for BM25 (see \Cref{tab:first-Stage}). Consequently, we select SPLADE-v3-DistilBERT as our primary first-stage retriever for the remainder of this work, as higher-recall learned sparse retriever increases the chance that the cross-encoders are evaluated on pools of passages with relevant candidates.

\subsection{Architectures}\label{sec:archis}

To gather as many insights as possible on the behavior of different training objectives across diverse backbones, we include a diverse range of  architectures in our benchmark.

\begin{table}[t]
\centering
\small
\caption{Configurations of the models used in this work.}\label{tab:model-configs}
\resizebox{\columnwidth}{!}{
\begin{tabular}{rrc}
\toprule
\multirow{15}{*}{\rotatebox{90}{\textbf{Mini} ($<$100M)}} 
 & Base  & \textbf{MiniLM}: \texttt{microsoft/MiniLM-L12-H384-uncased} \\
 & Tokenizer & WordPiece / uncased, \textbf{Ctx. Len.}=512 \\
 & Architecture & \textbf{Layers}=12, \textbf{Params}=33M, \textbf{Hidden}=384, \textbf{Heads}=12 \\
\cmidrule{2-3}
 & Base  & \textbf{Ettin-17}: \texttt{jhu-clsp/ettin-encoder-17m} \\
 & Tokenizer & BPE / uncased, \textbf{Ctx. Len.}=8192 \\
 & Architecture & \textbf{Layers}=7, \textbf{Params}=17M, \textbf{Hidden}=256, \textbf{Heads}=4 \\
\cmidrule{2-3}
 & Base  & \textbf{Ettin-32}: \texttt{jhu-clsp/ettin-encoder-32m} \\
 & Tokenizer & BPE / uncased, \textbf{Ctx. Len.}=8192 \\
 & Architecture & \textbf{Layers}=10, \textbf{Params}=32M, \textbf{Hidden}=384, \textbf{Heads}=6 \\
\cmidrule{2-3}
 & Base  & \textbf{Ettin-68}: \texttt{jhu-clsp/ettin-encoder-68m} \\
 & Tokenizer & BPE / uncased, \textbf{Ctx. Len.}=8192 \\
 & Architecture & \textbf{Layers}=19, \textbf{Params}=68M, \textbf{Hidden}=512, \textbf{Heads}=8 \\
\midrule
\multirow{18}{*}{\rotatebox{90}{\textbf{Base}}} 
 & Base  & \textbf{BERT}: \texttt{google-bert/bert-base-uncased} \\
 & Tokenizer & WordPiece / uncased, \textbf{Ctx. Len.}=512 \\
 & Architecture & \textbf{Layers}=12, \textbf{Params}=110M, \textbf{Hidden}=768, \textbf{Heads}=12 \\
\cmidrule{2-3}
 & Base  & \textbf{RoBERTa}: \texttt{FacebookAI/roberta-base} \\
 & Tokenizer & BPE / cased, \textbf{Ctx. Len.}=512 \\
 & Architecture & \textbf{Layers}=12, \textbf{Params}=125M, \textbf{Hidden}=768, \textbf{Heads}=12 \\
\cmidrule{2-3}
 & Base  & \textbf{ELECTRA}: \texttt{google/electra-base-discriminator} \\
 & Tokenizer & WordPiece / uncased, \textbf{Ctx. Len.}=512 \\
 & Architecture & \textbf{Layers}=12, \textbf{Params}=110M, \textbf{Hidden}=768, \textbf{Heads}=12 \\
\cmidrule{2-3}
 & Base  & \textbf{DeBERTaV3}: \texttt{microsoft/deberta-v3-base} \\
 & Tokenizer & BPE / cased, \textbf{Ctx. Len.}=512 \\
 & Architecture & \textbf{Layers}=12, \textbf{Params}=184M, \textbf{Hidden}=768, \textbf{Heads}=12 \\
\cmidrule{2-3}
 & Base  & \textbf{Ettin-150}: \texttt{jhu-clsp/ettin-encoder-150m} \\
 & Tokenizer & BPE / uncased, \textbf{Ctx. Len.}=8192 \\
 & Architecture & \textbf{Layers}=22, \textbf{Params}=150M, \textbf{Hidden}=768, \textbf{Heads}=12 \\
\bottomrule
\end{tabular}
}
\end{table}

\begin{enumerate}
    \item BERT \cite{bert}: The first transformer-based model, pre-trained using masked language modeling (MLM) and next sentence prediction (NSP) on English Wikipedia and BookCorpus.
    \item MiniLM-v2 \cite{minilmv2}: A highly efficient version of BERT for low-latency applications, but pre-trained using a generalization of the deep self-attention distillation approach \cite{minilm}.
    \item RoBERTa \cite{liu2019roberta}: A robustly optimized version of BERT, with an improved pre-training, including more data and the removal of the NSP task.
    \item ELECTRA \cite{clark2020electra}: Which uses an efficient discriminative pre-training objective (replaced token detection), with recent successes in ranking \cite{schlattRankDistiLLMClosingEffectiveness2025}.
    \item DeBERTaV3 \cite{he2021debertav3}: A model utilizing disentangled attention and improved masked-language modelling that has proven very effective for re-ranking \cite{lassance2023naverlabseuropesplade}.
    \item Ettin \cite{ettin}: The Ettin suite is a comprehensive suite of encoder-only models, scaling from 17M to 1B parameters, based on ModernBERT~\cite{modernbert}, recently proposed as an improvement to BERT, with new architectural traits such as RoPE \cite{roformer} and sliding-window attention.
\end{enumerate}

Note that we purposefully restrict our scope to encoder-only backbones, as these models have been shown to outperform decoder-only alternatives for discriminative tasks such as classification and ranking \cite{ettin, Dejean2024ATC}.
All model configurations are summarized in \Cref{tab:model-configs}. We now describe the training objectives evaluated across these backbones.

\subsection{Training Strategies}\label{sec:training_strategies}

In this section, we detail the six training strategies compared in this work, all of which have proven effective for optimizing cross-encoders.  All models are fine-tuned using \CLS{} pooling after the final transformer layer \cite{monobert}. Our analysis compares (i) supervised objectives derived from raw MS MARCO \cite{msmarco} labels, including the choice of the negative sampling strategy, and (ii) distillation objectives introduced by \citet{hofstatterImprovingEfficientNeural2020} and \citet{schlattRankDistiLLMClosingEffectiveness2025}.

\subsubsection{Supervised Objectives}

First, the following losses train directly on MS MARCO \cite{msmarco} binary relevance labels.\\

\textbf{Binary Cross-Entropy (BCE)}
Following monoBERT~\cite{monobert}, this objective views the ranking task as a pointwise classification. We use the BCE objective in a \textit{pairwise sampling} setup, where each query is associated with a pair of passages: one positive $(p_+, s_+)$ and one negative $(p_-, s_-)$.
$$L_{\text{BCE}} = - \log(\sigma (s_+)) - \log\left(1-\sigma
(s_-)\right)$$

\noindent with $\sigma$ the sigmoid function.\\

\textbf{Hinge Loss} As used in CEDR \cite{macavaney2019ContextualWR},
this pairwise objective enforces a margin of at least 1 between the model's scores for the positive ($s_+$) and
negative ($s_- $) documents:
$$\mathcal L_{\text{Hinge}} = \max(0\ , 1 - (s_+ - s_- ))$$

We use the same random triplet $(q, p_+, p_-)$ sampling for both BCE and Hinge losses, ensuring fair comparison.\\

\textbf{InfoNCE} This listwise objective is functionally equivalent to ListNet \cite{caoLearningRankPairwise2007} and the LCE loss \cite{pradeepSqueezingWaterStone2022}, using binary relevance labels $y_i$:
$$\mathcal L_{\text{InfoNCE}} = - \sum_{i=1}^n y_i \log
\left(\frac{e^{s_i}}{\sum_{j=1}^n e^{s_j}} \right)$$

To sample the negatives, we replicate the training strategy of \citet{schlattRankDistiLLMClosingEffectiveness2025} and randomly sample 7 negatives from the pool of top-500 passages retrieved by ColBERTv2 \cite{santhanam2022colbertv2} for each of the 10k queries in the dataset they released.

\subsubsection{Distillation Objectives}
The following objectives replace binary labels with soft scores from a teacher model \cite{hinton2015distillingknowledgeneuralnetwork}, providing a more fine-grained supervision signal and overcoming the known issue with false negatives in MS~MARCO \cite{thakurHardNegativesHard2025,qu-etal-2021-rocketqa}.\\

\textbf{MarginMSE}
As proposed by \citet{hofstatterImprovingEfficientNeural2020}, this pairwise objective regresses the student's margin $s_{+} - s_-$ to match the teacher's margin $s_{+}^t - s_-^t$ instead of fitting the logit values:
$$\mathcal L_{\text{MarginMSE}} = \text{MSE}(\ s_{+}^t - s_-^t,\ s_{+} - s_-\ )$$

The teacher is an ensemble of three cross-encoders---BERT-base \cite{bert}, BERT-large, and ALBERT \cite{lan2020albertlitebertselfsupervised}, each trained with the RankNet loss \cite{burgesLearningRankUsing2005}---used to annotate the top-1000 BM25 passages for the queries of MS~MARCO. We use the original triplets and teacher scores released by the authors to ensure exact reproducibility.\\

The last two objectives use a strong LLM teacher model, i.e., RankZephyr \cite{pradeep2023rankzephyr},
to rank a list of passages $p_1 > \ldots > p_n$.\\

\textbf{DistillRankNet}\label{strat:DistillRankNet} 
Following \citet{schlattRankDistiLLMClosingEffectiveness2025}, this objective is a teacher-student adaptation of RankNet \cite{burgesLearningRankUsing2005} that considers all the pairs formed by each passage ranked higher in the teacher's ranking list, and those ranked below:
$$\mathcal L_{\text{DistillRankNet}} = -\sum_{i=1}^n\sum_{j=i+1}^n \log
\sigma(s_i - s_j)$$

\textbf{ADR-MSE (Adaptive Distillation Rank MSE)} \label{strat:ADR-MSE} Also introduced in \cite{schlattRankDistiLLMClosingEffectiveness2025}, applies a weighted MSE to the scores based on their rank in the teacher's ranking list, emphasizing the top-tier results:
$$\mathcal{L}_{\text{ADR-MSE}} = \frac{1}{n} \sum_{i=1}^{n} \frac{1}{\log_2(i +
1)} \left( i - \mathbf{r}_i^s  \right)^2$$

Using a differentiable approximation of the student rank $\mathbf{r}_i^s$ of passage $p_i$, for a given temperature $T$ (usually $T=1$ \cite{schlattRankDistiLLMClosingEffectiveness2025}):
$$\mathbf{r}_i^s = 1 + \sum_{j \neq i} \sigma\!\left(\frac{s_j -
s_i}{T}\right)$$

For both $\mathcal L_{\text{DistillRankNet}}$ and
$\mathcal{L}_{\text{ADR-MSE}}$, we use re-ranking signals from RankZephyr \cite{pradeep2023rankzephyr} of the top 50 passages retrieved by ColBERTv2 for a subset of 10k queries from MS MARCO,  shared by \citet{schlattRankDistiLLMClosingEffectiveness2025}, as their results suggest this depth is the most suitable to fine-tune cross-encoders.

\subsection{Implementation Details} The training pipeline is based on PyTorch \cite{pytorch} and \texttt{transformers}\footnote{https://github.com/huggingface/transformers}. Each training run is conducted using a single NVIDIA H100 (80GB) GPU, with all experiments lasting less than 20 hours (around 2h30 for training \textit{Mini} backbones). Across training strategies, we normalize each batch by the total number of documents to keep the number of samples comparable. This implies that each batch for the DistillRankNet and ADR-MSE losses is composed of a single query, along with its ranked list of 50 passages. 

As tokenization can be a major source of discrepancy given the various context length limit for the backbones considered in this work (e.g in \Cref{tab:model-configs}, 512 for most BERT-like backbones, vs 8192 for ModernBERT), we keep all input tokenization, maximum sequence lengths, and evaluation scripts constant across experiments to ensure fair comparisons. Notably, we follow \citet{schlattRankDistiLLMClosingEffectiveness2025} and truncate all queries and passages respectively to 32 and 256 tokens.

\subsection{Training Hyperparameters}\label{subsec:Base}

To maintain a controlled environment across our diverse backbones and loss functions, we establish a standardized training protocol designed to isolate the impact of architecture and objective function. This framework ensures consistency by fixing several key parameters across all runs:


\textbf{Evaluation Candidate Generation:} We maintain a fixed candidate set size of n=1000 passages per query, retrieved using the SPLADE-v3-DistilBERT first-stage retriever.

\textbf{Optimization Framework}: Fine-tuning is performed using the AdamW optimizer with $\epsilon=1\cdot10^{-8}$. We adapt batch sizes and learning rates per model size class via the preliminary experiments described in \Cref{sec:hyperparams}, but keep the overall scheduling uniform. We perform 100 steps per epoch (max. of 200k optimization steps per training) with a linear warmup for the first 5k steps.

\textbf{Validation Strategy} During training, we perform validation every 50 epochs to monitor convergence. Given the computational cost of validating on complete datasets, but considering the fact that cross-encoders are prone 
to overfitting on the training corpus, we follow the \texttt{sentence-transformers} library guidelines \footnote{\href{https://sbert.net/index.html}{https://sbert.net/index.html}} and use nano-BEIR \cite{NanoBEIRMultilingualInformation2026} as a proxy for generalization effectiveness. These subsets of the 13 public datasets of BEIR, allow for frequent validation without the overhead of using the full datasets.
Note that because we use nano-BEIR to find our best checkpoints (yet do not directly train on it), we consider BEIR as a \textit{semi OOD} benchmark, and only regard the LoTTE and Robust04 datasets as truly OOD.

\begin{table}[t]
\centering
\caption{Hyperparameter grid search results on proxy models. Values represent the mean nDCG@10 $\pm$ standard deviation across three random seeds. Bold values indicate the best learning rate ($\eta$) and batch size (b\_size). Total training time on an NVIDIA H100 is provided for reproducibility.}
\label{tab:grid_search_results}
\resizebox{\columnwidth}{!}{%
\begin{tabular}{r|cc|cc}
\toprule
 & \multicolumn{2}{c|}{\textbf{MiniLM}} & \multicolumn{2}{c}{\textbf{BERT-base}} \\
$\eta$ \textbackslash \ b\_size & \textbf{32} & \textbf{64} & \textbf{32} & \textbf{64} \\
\midrule
$7 \times 10^{-5}$ & $58.47 \pm 0.12$ & $57.56 \pm 0.19$ & $56.63 \pm 0.15$ & $57.75 \pm 0.08$ \\
$1 \times 10^{-5}$ & $60.63 \pm 0.05$ & $60.73 \pm 0.07$ & \cellcolor{green!20}$\mathbf{60.38 \pm 0.11}$ & $\mathbf{60.30 \pm 0.11}$ \\
$7 \times 10^{-6}$ & $60.33 \pm 0.05$ & \cellcolor{green!20}$\mathbf{60.95 \pm 0.03}$ & $60.05 \pm 0.08$ & $59.90 \pm 0.03$ \\
$3 \times 10^{-6}$ & $60.41 \pm 0.03$ & $\mathbf{60.96 \pm 0.05}$ & $59.97 \pm 0.04$ & $59.70 \pm 0.05$ \\
\midrule
Duration & 2h30 & 5h & 4h & 8h \\
\bottomrule
\end{tabular}%
}
\vspace{-.3cm}
\end{table}

\subsection{Phased Experimental Strategy}\label{sec:hyperparams}
    
Given the combinatorial explosion inherent in testing every experimental parameter — encompassing nine backbones, six loss functions, and multiple hyperparameter variants — we adopt a phased strategy for hyperparameter search. This approach allows us to identify the best training settings in a controlled environment, without requiring an exhaustive and computationally prohibitive grid search across all configurations.\\

\textbf{Phase I: Selection of learning rates and batch sizes.} Following the hypothesis that optimal hyperparameters depend primarily on model scale rather than architecture or training objective \cite{kaplan2020scaling}, we run a preliminary grid search over batch size and learning rate $\eta$ on two proxy models with the BCE loss $\mathcal{L_\text{BCE}}$: MiniLM-L12-v2 for \textit{Mini} models ($<$100M parameters) and BERT-base for \textit{Base} models (100M–400M parameters). This search yields a baseline configuration per size class, later extended to other loss functions. Our findings reveal distinct best configurations tailored to model's capacity, as
detailed in Table \ref{tab:grid_search_results}.

\begin{enumerate}
    \item \textbf{\textit{Base} Models} For BERT-scale architectures, a learning rate of $\eta = 1 \times 10^{-5}$ consistently yields the highest effectiveness. As setting the batch size to 64 provides no significant effectiveness gain while nearly doubling the required training time, we fix the batch size at 32.
    \item \textbf{\textit{Mini} Models} For MiniLM and the three smallest models of the Ettin suite, a lower learning rate of $\eta=7 \times 10^{-6}$ combined with a larger batch size of 64 provides the most stable convergence: while $\eta=3 \times 10^{-6}$ achieved similar mean effectiveness, $7 \times 10^{-6}$ yields lower variance across seeds, making it a more robust choice for reproduction.
\end{enumerate}

These selected hyperparameters are subsequently applied to all backbones within their respective size classes to ensure a fair and optimized comparison in the final benchmark. \\

\textbf{Phase II: Benchmarking Backbones and Objectives}\label{sec:Backbones&Losses} With the optimization parameters fixed, we proceed to the full-scale benchmark. This phase evaluates the intersection of architecture and training objectives, totalling 162 unique experimental runs (9 backbones $\times$ 6 losses $\times$ 3 seeds). This systematic execution allows us to isolate the relative contribution of the \textit{training strategy} versus the \textit{backbone architecture} to the model's final ranking effectiveness.
\begin{table}[t]
\centering
\caption{Computational budget and experimental breakdown of our phased approach to identify optimal hyperparameters. $N_{H}$ is the number of hyperparameter sets , $N_{B}$ the number of backbones, $N_{L}$ the number of losses and $N_D$ the number of evaluation datasets (3 corresponds to ID-only evaluations, 23 to the complete suite).}
\label{tab:comp_budget}
\resizebox{\columnwidth}{!}{%
\begin{tabular}{@{}lccccc|cc@{}}
\toprule
\textbf{Phase} & $N_{H}$ & $N_{B}$ & $N_{L}$ & seeds & \textbf{\#Runs} & $N_{D}$ & \textbf{\#Evals} \\ \midrule
\textbf{I: Hyperparameters} & 8 & 2 & 1 & 3 & 48 & 3 & 144\\
\textbf{II: Backbones / Objectives}    & 1 & 9 & 6 & 3 & 162 & 23 & 3726\\
\textbf{\ref{sec:Negatives_comp}: Losses / Negatives}    & 1 & 3 & 3 & 3 & 27 & 23 & 621\\
\midrule
\textbf{Total} & & & & & \textbf{237} & & \textbf{4491}\\ \bottomrule
\end{tabular}%
}
\vspace{-.3cm}
\end{table}

\subsection{Reproducibility Statement}
A key tenet of our methodology is the mitigation of initialization bias. Every unique experimental configuration is executed with three fixed random seeds. To ensure a faithful comparison of the tested parameters and architectures, we report the mean effectiveness along with the standard deviation across the seeds.

Every training experiment is defined by a standalone YAML configuration file specifying all hyperparameters, model architectures and evaluation datasets. We release these configuration files alongside our new unified codebase, allowing researchers to reproduce our exact results with one command or use them as templates for benchmarking new training strategies in a comparable environment.
Finally, we will release the best-performing model checkpoints for each training strategy on HuggingFace upon publication, to serve as a fully reproducible reference point for future works on cross-encoder training and distillation.

\section{Results and Analysis}
In this section, we present the full benchmark and analyze how training objectives, backbones, and negative sampling shape cross-encoder effectiveness in our controlled setup.

\begin{figure*}[t]
    \centering
    \vspace{-.8cm}
    \begin{subfigure}{\textwidth}
        \centering
        \caption{ID evaluation}
        \includegraphics[width=\textwidth]{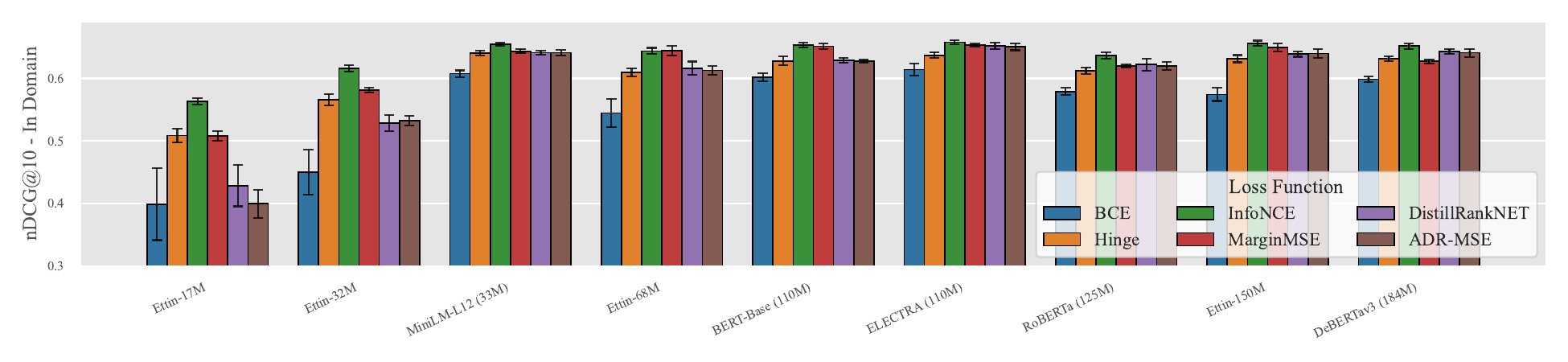}
    \end{subfigure}
    \vspace{-.4cm}
    \begin{subfigure}{\textwidth}
        \centering
        \caption{Evaluation on BEIR-13 (semi OOD in our setup)}
        \includegraphics[width=\textwidth]{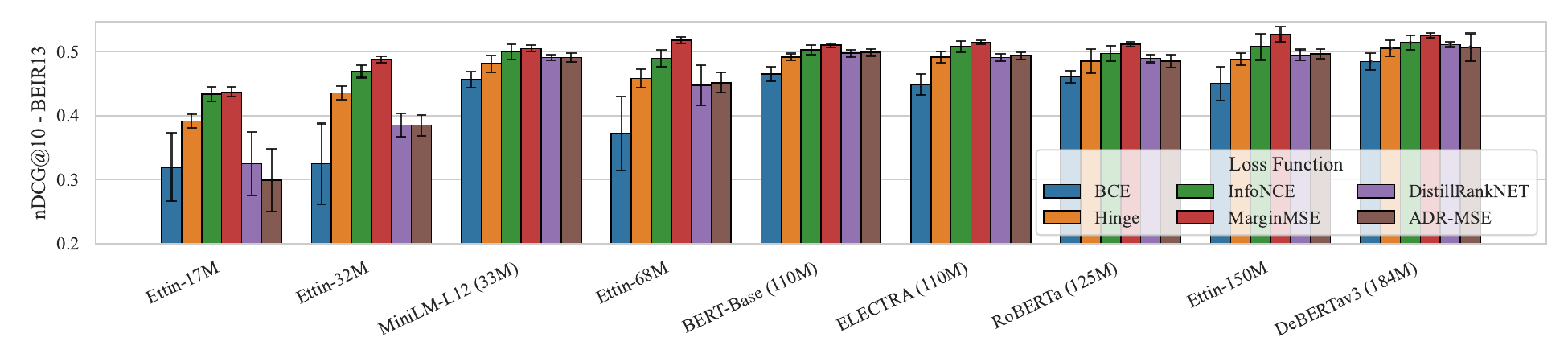}
    \end{subfigure}
    \vspace{-.4cm}
    \begin{subfigure}{\textwidth}
        \centering
        \caption{Evaluation on OOD (LoTTe Search + Robust04)}
        \includegraphics[width=\textwidth]{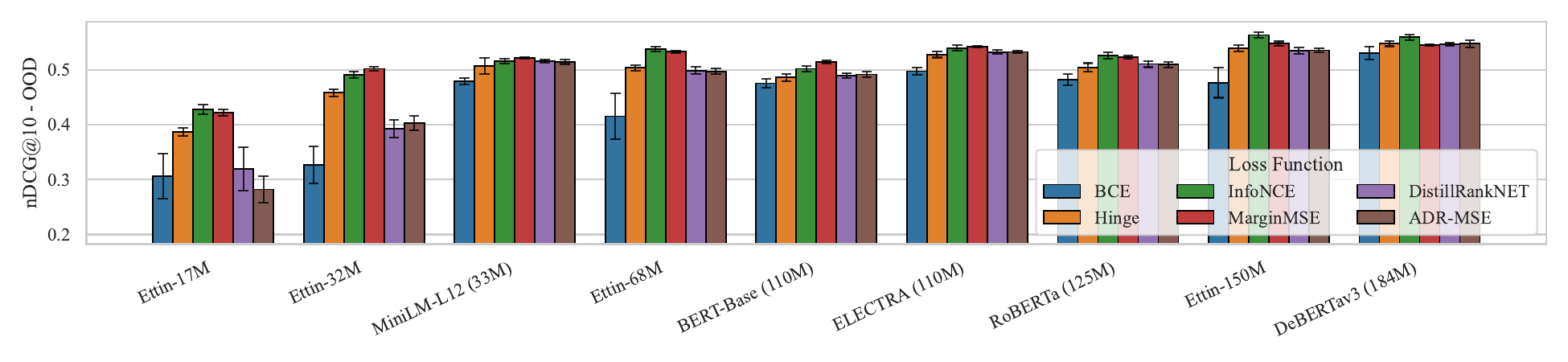}
    \end{subfigure}
    \vspace{-.4cm}
    \caption{Comprehensive evaluation of cross-encoder backbones across six training strategies. We report the mean nDCG@10 across three random seeds for in-domain (ID), BEIR-13 (semi-OOD), and the combined OOD suite (LoTTe Search and Robust04). Error bars indicate the standard deviation across runs, illustrating the relative stability of different distillation strategies.}
    \label{fig:backbonesXlosses}
\end{figure*}

\subsection{Full Results}

We present aggregated results in \Cref{fig:backbonesXlosses}, which summarizes the effectiveness of nine encoder-only backbones across six training objectives, and detail all values in \Cref{tab:results}. We report the mean nDCG@10 over three random seeds, evaluated across three distinct scenarios: ID (MS~MARCO dev, TREC-DL'19, TREC-DL'20), the BEIR-13 benchmark (\emph{semi-OOD}, since we use nano-BEIR to select checkpoints), which includes the following list of datasets: ArguAna, Climate-FEVER, DBPedia, FEVER, FiQA, HotpotQA, NFCorpus, Natural Questions, Quora Question Pairs, SCIDOCS, SciFact, Touché-2020 (v2), and TREC-COVID. Finally, we also include a true OOD suite, comprising LoTTE Search and Robust04. With only three seeds, differences between close-performing objectives should be interpreted with caution; our analysis focuses on consistent trends that hold across backbones and evaluation settings.

\begin{table}[t]
\centering
\small
\caption{Reproducing the results of \citet{schlattRankDistiLLMClosingEffectiveness2025} with ELECTRA. We report nDCG@10 on TREC DL 19 and 20 from the original paper, along with our own reproduction. 
\;\;\textcolor{OIOrange}{\rule{1.2ex}{1.2ex}} Results from \citet{schlattRankDistiLLMClosingEffectiveness2025}
\;\;\textcolor{OISkyBlue}{\rule{1.2ex}{1.2ex}} Our framework} \label{tab:rankdistill}
\vspace{-.3cm}
\resizebox{\columnwidth}{!}{%
\begin{tabular}{@{}rlc|cc@{}}
    \toprule
    \textbf{First Stage} & \textbf{Strategy} & \textbf{Distillation} & \textbf{DL 19} & \textbf{DL 20} \\ 
    \midrule
    ColBERTv2 & First Stage & -- & $73.2$ & $72.4$ \\
    SPLADE-v3 & First Stage & -- & $75.2$ & $74.2$ \\
    \midrule

    \rowcolor{SchlattRow}
    \multicolumn{5}{c}{\textit{Results from \citet{schlattRankDistiLLMClosingEffectiveness2025}}} \\ 
    \rowcolor{SchlattRow}
    \midrule
    \rowcolor{SchlattRow}
    ColBERTv2 (top100)  & InfoNCE (ColBERTv2 Neg) &  -            & $73.9$ & $76.0$ \\
    \rowcolor{SchlattRow}
    ColBERTv2 (top100)  & DistillRankNet          & Rank-DistiLLM & $77.4$ & $75.4$ \\ 
    \midrule

    \rowcolor{OurRow}
    \multicolumn{5}{c}{\textit{Our Framework}} \\ 
    \rowcolor{OurRow}
    \midrule
    \rowcolor{OurRow}
    SPLADE-v3 (top100) & InfoNCE (ColBERTv2 Neg) & -      & $75.8 \pm 0.02$ & $74.6 \pm 0.04$\\
    \rowcolor{OurRow}
    SPLADE-v3 (top100) & DistillRankNet & Rank-DistiLLM & $77.4 \pm 0.09$ & $75.1 \pm 0.03$ \\
    \rowcolor{OurRow}
    \midrule
    \rowcolor{OurRow}
    SPLADE-v3 (top1k) & InfoNCE (ColBERTv2 Neg) &  -            & $75.2 \pm 0.03$ & $74.5 \pm 0.05$ \\
    \rowcolor{OurRow}
    SPLADE-v3 (top1k) & DistillRankNet          & Rank-DistiLLM & $77.0 \pm 0.08$ & $74.4 \pm 0.04$ \\
    \bottomrule
\end{tabular}
}
\vspace{-.3cm}
\end{table}

\textbf{Validation of our methodology.} We first validate our implementation against \citet{schlattRankDistiLLMClosingEffectiveness2025}. This comparison allows us to confirm the correctness of our training pipeline before extending the analysis to the full extent of this study. 

Table~\ref{tab:rankdistill} compares the TREC-DL'19 and TREC-DL'20 effectiveness reported by \citet{schlattRankDistiLLMClosingEffectiveness2025} for ELECTRA with our unified evaluation setup in this paper. It also validates our use of SPLADE-v3-DistilBERT for the first-stage, compared to ColBERTv2 in the original study. Our reproduced DistillRankNet results on TREC-DL'19 perfectly recover the effectiveness reported by \citet{schlattRankDistiLLMClosingEffectiveness2025} when using an equivalent top-100 pool ($77.4\pm0.09$ vs. $77.4$ nDCG@10). On TREC-DL'20, our reproduction reaches $75.1\pm0.03$, narrowing the gap to the original $75.4$. These results indicate that our implementation successfully recovers the effectiveness of the Rank-DistiLLM objectives.

\textbf{Statistical Analysis.}\label{sec:stat_methodology} We assess significance using a Friedman test with a Nemenyi post-hoc test \cite{demsarStatisticalComparisonsClassifiers2006}. We aggregate the per-dataset evaluations into the three scenarios above (ID, BEIR-13, OOD) and apply the Friedman test to rank the variable of interest (objectives in \Cref{sec:strategy_impact}, backbones in \Cref{sec:backbone_and_scaling}) across the resulting combinations; Nemenyi then identifies which pairs differ significantly. These non-parametric, rank-based tests avoid normality assumptions and handle benchmark heterogeneity better than ANOVA, since they compare methods on ranks rather than raw scores.
However, a non-significant Nemenyi result does not by itself imply two methods are equivalent—it may simply reflect low statistical power. To distinguish these cases, we complement it with a TOST (Two One-Sided Tests) equivalence test~\cite{Schuirmann1987ACO} on the paired per-instance scores. TOST reverses the usual burden of proof: for a pre-specified equivalence margin $\epsilon$, its null hypothesis is \emph{non}-equivalence, $|\mu_1-\mu_2|\geq\epsilon$, assessed through two one-sided $t$-tests of $\mu_1-\mu_2\leq-\epsilon$ and of $\mu_1-\mu_2\geq\epsilon$. Only when both are rejected at level $\alpha$ can we positively conclude that the difference between the two methods' mean scores lies within $\pm\epsilon$. We set $\epsilon=0.05$ nDCG@10 and $\alpha=0.05$.\\

\begin{table*}
\caption{Mean nDCG@10 (3 seeds) across all benchmarks using the SPLADE-v3 top-1000 pipeline. \textbf{Bold} is the best overall result per column; \textit{italics} indicate the best objective per backbone. Cell shading per column: \colorbox{red!15}{red}~=~above median, \colorbox{blue!15}{blue}~=~below.} \label{tab:results}
\vspace{-.3cm}
\resizebox{0.77\textwidth}{!}{%
\begin{tabular}{llcccccc}
\toprule
 & & \multicolumn{3}{c}{\textbf{ID}} & & \multicolumn{2}{c}{\textbf{OOD}} \\
 \cmidrule(lr){3-5} \cmidrule(lr){7-8}
\textbf{Backbone model} & \textbf{Loss (Negatives or Distill data)} & MSM & DL19 & DL20 & BEIR-13 & Lotte-S & Robust \\
\midrule
\multirow[c]{6}{*}{\textbf{Ettin-17M}} & BCE & \cellcolor{blue!18}0.265 & \cellcolor{blue!20}0.467 & \cellcolor{blue!20}0.464 & \cellcolor{blue!18}0.319 & \cellcolor{blue!18}0.323 & \cellcolor{blue!18}0.225 \\
 & Hinge & \cellcolor{blue!10}0.347 & \cellcolor{blue!11}0.584 & \cellcolor{blue!9}0.595 & \cellcolor{blue!10}0.391 & \cellcolor{blue!11}0.400 & \cellcolor{blue!11}0.322 \\
 & InfoNCE (ColBERTv2) & \cellcolor{blue!5}0.394 & \cellcolor{blue!7}0.639 & \cellcolor{blue!4}0.657 & \cellcolor{blue!6}0.433 & \cellcolor{blue!7}0.443 & \cellcolor{blue!8}0.354 \\
 & MarginMSE (\citet{hofstatterImprovingEfficientNeural2020}) & \cellcolor{blue!8}0.360 & \cellcolor{blue!11}0.580 & \cellcolor{blue!10}0.584 & \cellcolor{blue!6}0.437 & \cellcolor{blue!7}0.442 & \cellcolor{blue!11}0.320 \\
 & DistillRankNET (RankDistiLLM \cite{schlattRankDistiLLMClosingEffectiveness2025}) & \cellcolor{blue!18}0.271 & \cellcolor{blue!16}0.523 & \cellcolor{blue!18}0.491 & \cellcolor{blue!17}0.325 & \cellcolor{blue!17}0.332 & \cellcolor{blue!16}0.256 \\
 & ADR-MSE (RankDistiLLM \cite{schlattRankDistiLLMClosingEffectiveness2025}) & \cellcolor{blue!20}0.248 & \cellcolor{blue!19}0.474 & \cellcolor{blue!19}0.477 & \cellcolor{blue!20}0.299 & \cellcolor{blue!20}0.296 & \cellcolor{blue!19}0.213 \\
\midrule
\multirow[c]{6}{*}{\textbf{Ettin-32M}} & BCE & \cellcolor{blue!12}0.326 & \cellcolor{blue!15}0.530 & \cellcolor{blue!17}0.495 & \cellcolor{blue!17}0.325 & \cellcolor{blue!15}0.352 & \cellcolor{blue!20}0.201 \\
 & Hinge & \cellcolor{blue!4}0.399 & \cellcolor{blue!6}0.654 & \cellcolor{blue!5}0.644 & \cellcolor{blue!6}0.435 & \cellcolor{blue!4}0.475 & \cellcolor{blue!7}0.374 \\
 & InfoNCE (ColBERTv2) & \cellcolor{red!1}0.439 & 0.723 & \cellcolor{blue!2}0.686 & \cellcolor{blue!2}0.469 & \cellcolor{blue!1}0.507 & \cellcolor{blue!4}0.410 \\
 & MarginMSE (\citet{hofstatterImprovingEfficientNeural2020}) & \cellcolor{blue!1}0.426 & \cellcolor{blue!5}0.665 & \cellcolor{blue!5}0.653 & 0.488 & 0.516 & \cellcolor{blue!2}0.429 \\
 & DistillRankNET (RankDistiLLM \cite{schlattRankDistiLLMClosingEffectiveness2025}) & \cellcolor{blue!10}0.346 & \cellcolor{blue!9}0.604 & \cellcolor{blue!6}0.636 & \cellcolor{blue!11}0.385 & \cellcolor{blue!9}0.414 & \cellcolor{blue!13}0.287 \\
 & ADR-MSE (RankDistiLLM \cite{schlattRankDistiLLMClosingEffectiveness2025}) & \cellcolor{blue!9}0.351 & \cellcolor{blue!9}0.609 & \cellcolor{blue!6}0.635 & \cellcolor{blue!11}0.385 & \cellcolor{blue!9}0.423 & \cellcolor{blue!12}0.302 \\
\midrule
\multirow[c]{6}{*}{\textbf{MiniLM-L12 (33M)}} & BCE & \cellcolor{red!1}0.441 & \cellcolor{blue!3}0.684 & \cellcolor{blue!1}0.698 & \cellcolor{blue!4}0.456 & \cellcolor{blue!3}0.486 & \cellcolor{blue!1}0.442 \\
 & Hinge & \cellcolor{red!6}0.450 & \cellcolor{red!4}0.736 & \cellcolor{red!14}0.735 & \cellcolor{blue!1}0.481 & 0.513 & \cellcolor{red!5}0.476 \\
 & InfoNCE (ColBERTv2) & \cellcolor{red!15}0.469 & \cellcolor{red!14}0.758 & \cellcolor{red!15}0.737 & \cellcolor{red!6}0.500 & \cellcolor{red!1}0.521 & \cellcolor{red!10}0.489 \\
 & MarginMSE (\citet{hofstatterImprovingEfficientNeural2020}) & \cellcolor{red!13}0.465 & \cellcolor{red!4}0.736 & \cellcolor{red!11}0.730 & \cellcolor{red!8}0.505 & \cellcolor{red!2}0.525 & \cellcolor{red!14}0.504 \\
 & DistillRankNET (RankDistiLLM \cite{schlattRankDistiLLMClosingEffectiveness2025}) & 0.438 & \cellcolor{red!8}0.745 & \cellcolor{red!17}0.740 & \cellcolor{red!1}0.490 & 0.515 & \cellcolor{red!20}\textbf{0.520} \\
 & ADR-MSE (RankDistiLLM \cite{schlattRankDistiLLMClosingEffectiveness2025}) & 0.438 & \cellcolor{red!9}0.746 & \cellcolor{red!16}0.740 & \cellcolor{red!1}0.491 & 0.514 & \cellcolor{red!19}0.517 \\
\midrule
\multirow[c]{6}{*}{\textbf{Ettin-68M}} & BCE & \cellcolor{blue!6}0.384 & \cellcolor{blue!7}0.633 & \cellcolor{blue!8}0.617 & \cellcolor{blue!12}0.372 & \cellcolor{blue!6}0.446 & \cellcolor{blue!15}0.265 \\
 & Hinge & \cellcolor{blue!1}0.432 & \cellcolor{blue!2}0.708 & \cellcolor{blue!2}0.689 & \cellcolor{blue!3}0.458 & \cellcolor{red!2}0.523 & \cellcolor{blue!4}0.408 \\
 & InfoNCE (ColBERTv2) & \cellcolor{red!14}0.466 & \cellcolor{red!4}0.736 & \cellcolor{red!11}0.730 & 0.490 & \cellcolor{red!11}0.553 & 0.456 \\
 & MarginMSE (\citet{hofstatterImprovingEfficientNeural2020}) & \cellcolor{red!14}0.466 & \cellcolor{red!5}0.738 & \cellcolor{red!10}0.728 & \cellcolor{red!15}0.518 & \cellcolor{red!9}0.545 & \cellcolor{red!4}0.472 \\
 & DistillRankNET (RankDistiLLM \cite{schlattRankDistiLLMClosingEffectiveness2025}) & \cellcolor{blue!3}0.407 & \cellcolor{red!4}0.736 & 0.706 & \cellcolor{blue!4}0.447 & 0.512 & \cellcolor{blue!2}0.429 \\
 & ADR-MSE (RankDistiLLM \cite{schlattRankDistiLLMClosingEffectiveness2025}) & \cellcolor{blue!3}0.409 & \cellcolor{red!4}0.735 & \cellcolor{blue!1}0.695 & \cellcolor{blue!4}0.451 & 0.513 & \cellcolor{blue!3}0.419 \\
\midrule
\multirow[c]{6}{*}{\textbf{BERT-Base (110M)}} & BCE & 0.434 & \cellcolor{blue!4}0.681 & \cellcolor{blue!2}0.691 & \cellcolor{blue!3}0.465 & \cellcolor{blue!3}0.485 & \cellcolor{blue!3}0.426 \\
 & Hinge & \cellcolor{red!3}0.444 & 0.722 & \cellcolor{red!5}0.718 & \cellcolor{red!1}0.491 & \cellcolor{blue!2}0.490 & \cellcolor{red!1}0.464 \\
 & InfoNCE (ColBERTv2) & \cellcolor{red!13}0.465 & \cellcolor{red!15}0.759 & \cellcolor{red!15}0.737 & \cellcolor{red!7}0.503 & \cellcolor{blue!1}0.505 & \cellcolor{red!7}0.483 \\
 & MarginMSE (\citet{hofstatterImprovingEfficientNeural2020}) & \cellcolor{red!14}0.465 & \cellcolor{red!11}0.751 & \cellcolor{red!16}0.738 & \cellcolor{red!11}0.510 & \cellcolor{red!1}0.521 & \cellcolor{red!8}0.484 \\
 & DistillRankNET (RankDistiLLM \cite{schlattRankDistiLLMClosingEffectiveness2025}) & \cellcolor{blue!1}0.430 & \cellcolor{red!4}0.736 & \cellcolor{red!6}0.721 & \cellcolor{red!4}0.497 & \cellcolor{blue!2}0.493 & \cellcolor{red!4}0.472 \\
 & ADR-MSE (RankDistiLLM \cite{schlattRankDistiLLMClosingEffectiveness2025}) & \cellcolor{blue!1}0.427 & \cellcolor{red!5}0.737 & \cellcolor{red!5}0.719 & \cellcolor{red!5}0.499 & \cellcolor{blue!2}0.494 & \cellcolor{red!6}0.480 \\
\midrule
\multirow[c]{6}{*}{\textbf{ELECTRA (110M)}} & BCE & \cellcolor{red!5}0.447 & \cellcolor{blue!2}0.699 & \cellcolor{blue!1}0.696 & \cellcolor{blue!4}0.449 & 0.518 & \cellcolor{blue!5}0.394 \\
 & Hinge & \cellcolor{red!9}0.456 & 0.727 & \cellcolor{red!10}0.728 & \cellcolor{red!1}0.491 & \cellcolor{red!7}0.539 & \cellcolor{red!2}0.468 \\
 & InfoNCE (ColBERTv2) & \cellcolor{red!19}0.477 & \cellcolor{red!11}0.752 & \cellcolor{red!20}0.745 & \cellcolor{red!10}0.507 & \cellcolor{red!10}0.548 & \cellcolor{red!13}0.499 \\
 & MarginMSE (\citet{hofstatterImprovingEfficientNeural2020}) & \cellcolor{red!18}0.475 & \cellcolor{red!6}0.740 & \cellcolor{red!20}0.745 & \cellcolor{red!13}0.514 & \cellcolor{red!10}0.548 & \cellcolor{red!18}0.514 \\
 & DistillRankNET (RankDistiLLM \cite{schlattRankDistiLLMClosingEffectiveness2025}) & \cellcolor{red!2}0.441 & \cellcolor{red!20}\textbf{0.770} & \cellcolor{red!19}0.744 & \cellcolor{red!1}0.491 & \cellcolor{red!7}0.540 & \cellcolor{red!10}0.491 \\
 & ADR-MSE (RankDistiLLM \cite{schlattRankDistiLLMClosingEffectiveness2025}) & \cellcolor{red!2}0.441 & \cellcolor{red!18}0.765 & \cellcolor{red!20}\textbf{0.746} & \cellcolor{red!2}0.493 & \cellcolor{red!7}0.540 & \cellcolor{red!12}0.498 \\
\midrule
\multirow[c]{6}{*}{\textbf{RoBERTa (125M)}} & BCE & \cellcolor{blue!1}0.428 & \cellcolor{blue!5}0.665 & \cellcolor{blue!5}0.644 & \cellcolor{blue!3}0.461 & \cellcolor{blue!1}0.501 & \cellcolor{blue!6}0.385 \\
 & Hinge & 0.439 & \cellcolor{blue!2}0.702 & \cellcolor{blue!1}0.696 & 0.485 & \cellcolor{red!1}0.520 & \cellcolor{blue!3}0.427 \\
 & InfoNCE (ColBERTv2) & \cellcolor{red!12}0.461 & 0.729 & \cellcolor{red!6}0.721 & \cellcolor{red!4}0.497 & \cellcolor{red!6}0.537 & \cellcolor{red!3}0.470 \\
 & MarginMSE (\citet{hofstatterImprovingEfficientNeural2020}) & \cellcolor{red!9}0.456 & \cellcolor{blue!1}0.710 & \cellcolor{blue!1}0.694 & \cellcolor{red!12}0.511 & \cellcolor{red!7}0.539 & \cellcolor{blue!1}0.443 \\
 & DistillRankNET (RankDistiLLM \cite{schlattRankDistiLLMClosingEffectiveness2025}) & \cellcolor{blue!1}0.424 & 0.726 & \cellcolor{red!4}0.716 & 0.489 & 0.519 & \cellcolor{red!3}0.469 \\
 & ADR-MSE (RankDistiLLM \cite{schlattRankDistiLLMClosingEffectiveness2025}) & \cellcolor{blue!2}0.422 & 0.725 & \cellcolor{red!2}0.713 & 0.485 & 0.517 & \cellcolor{red!4}0.472 \\
\midrule
\multirow[c]{6}{*}{\textbf{Ettin-150M}} & BCE & \cellcolor{blue!2}0.420 & \cellcolor{blue!6}0.653 & \cellcolor{blue!5}0.650 & \cellcolor{blue!4}0.450 & \cellcolor{blue!1}0.509 & \cellcolor{blue!11}0.313 \\
 & Hinge & \cellcolor{red!5}0.447 & \cellcolor{red!7}0.743 & 0.705 & 0.488 & \cellcolor{red!12}0.556 & \cellcolor{blue!1}0.452 \\
 & InfoNCE (ColBERTv2) & \cellcolor{red!20}\textbf{0.478} & \cellcolor{red!13}0.754 & \cellcolor{red!15}0.736 & \cellcolor{red!10}0.507 & \cellcolor{red!20}\textbf{0.580} & \cellcolor{red!6}0.479 \\
 & MarginMSE (\citet{hofstatterImprovingEfficientNeural2020}) & \cellcolor{red!18}0.475 & \cellcolor{red!6}0.740 & \cellcolor{red!14}0.735 & \cellcolor{red!20}\textbf{0.527} & \cellcolor{red!14}0.560 & \cellcolor{red!9}0.487 \\
 & DistillRankNET (RankDistiLLM \cite{schlattRankDistiLLMClosingEffectiveness2025}) & \cellcolor{blue!1}0.430 & \cellcolor{red!15}0.760 & \cellcolor{red!10}0.727 & \cellcolor{red!3}0.495 & \cellcolor{red!8}0.543 & \cellcolor{red!12}0.496 \\
 & ADR-MSE (RankDistiLLM \cite{schlattRankDistiLLMClosingEffectiveness2025}) & \cellcolor{blue!1}0.429 & \cellcolor{red!13}0.756 & \cellcolor{red!14}0.736 & \cellcolor{red!4}0.496 & \cellcolor{red!8}0.543 & \cellcolor{red!13}0.499 \\
\midrule
\multirow[c]{6}{*}{\textbf{DeBERTav3 (184M)}} & BCE & 0.438 & \cellcolor{blue!5}0.668 & \cellcolor{blue!2}0.690 & \cellcolor{blue!1}0.484 & \cellcolor{red!10}0.550 & \cellcolor{blue!2}0.435 \\
 & Hinge & \cellcolor{red!6}0.451 & \cellcolor{red!8}0.745 & \cellcolor{blue!1}0.699 & \cellcolor{red!8}0.505 & \cellcolor{red!15}0.565 & 0.458 \\
 & InfoNCE (ColBERTv2) & \cellcolor{red!18}0.474 & \cellcolor{red!10}0.749 & \cellcolor{red!13}0.733 & \cellcolor{red!13}0.514 & \cellcolor{red!18}0.575 & \cellcolor{red!7}0.481 \\
 & MarginMSE (\citet{hofstatterImprovingEfficientNeural2020}) & \cellcolor{red!11}0.461 & 0.723 & \cellcolor{blue!1}0.699 & \cellcolor{red!19}0.525 & \cellcolor{red!14}0.560 & \cellcolor{red!3}0.471 \\
 & DistillRankNET (RankDistiLLM \cite{schlattRankDistiLLMClosingEffectiveness2025}) & 0.434 & \cellcolor{red!16}0.761 & \cellcolor{red!13}0.734 & \cellcolor{red!11}0.511 & \cellcolor{red!13}0.559 & \cellcolor{red!8}0.483 \\
 & ADR-MSE (RankDistiLLM \cite{schlattRankDistiLLMClosingEffectiveness2025}) & \cellcolor{blue!1}0.433 & \cellcolor{red!12}0.753 & \cellcolor{red!15}0.737 & \cellcolor{red!9}0.507 & \cellcolor{red!14}0.561 & \cellcolor{red!7}0.481 \\
\bottomrule
\end{tabular}
}
\end{table*}

\noindent Two primary observations emerge from this large-scale benchmark:
 
\textbf{Strategy Consistency} The choice of training objective exerts a consistent and often substantial impact on effectiveness, regardless of the specific backbone family employed (see \autoref{sec:strategy_impact}).

\textbf{Architecture-Objective Interaction} While scaling the backbone generally improves results, superior training objectives can effectively compensate for smaller model sizes. In our study, this includes InfoNCE and MarginMSE, as well as pairwise losses paired with hard negatives (see \autoref{sec:Negatives_comp}). This compensation effect is especially pronounced under domain shifts (see \autoref{sec:backbone_and_scaling}).

The following subsections provide a detailed analysis of these interactions, focusing on backbone scaling, objective robustness, and the trade-off between efficiency and effectiveness.

\subsection{Impact of Training Objective}\label{sec:strategy_impact}

Applying the methodology of \Cref{sec:stat_methodology} over the 27 (backbone, strategies) combinations confirms highly significant differences among objectives (Friedman $\chi^2=68.0$, $p<10^{-12}$). Nemenyi ($\alpha=0.05$, CD$=1.451$) reveals three statistically distinct tiers: a top group of InfoNCE (avg.\ rank 1.57) and MarginMSE (2.80), a middle tier of DistillRankNet (3.52), ADR-MSE (3.74), and Hinge (3.78), and BCE (5.59) at the bottom, significantly worse than every other objective. These rankings, as well as the TOST test ($p<10^{-19}$, $\epsilon=0.05$, $\alpha=0.05$), confirm the conclusion of \citet{schlattRankDistiLLMClosingEffectiveness2025} that the simpler DistillRankNet and ADR-MSE are statistically equivalent as distillation objectives. The results further position InfoNCE as a stronger option: its pairwise rank difference with both losses exceeds the Nemenyi critical difference.

\textit{BCE underperforms every other training objectives.} These statistical groupings align with a clear qualitative pattern visible in \Cref{fig:backbonesXlosses}: across virtually all backbones and evaluation settings, the pointwise BCE objective forms a lower bound, consistently dominated by objectives that explicitly compare candidates. This pattern is most pronounced on BEIR-13 and the OOD suite, suggesting that objectives emphasizing relative comparisons between candidates learn more transferable decision boundaries than independent binary classification.

\textit{Listwise LLM distillations do not significantly improve over InfoNCE or MarginMSE.} Within the top tier, TOST results ($p<10^{-8}$) confirm that MarginMSE---which relies on BM25-mined negatives with cross-encoder teacher scores---achieves statistically equivalent effectiveness to InfoNCE, which uses ColBERTv2 hard negatives (see \Cref{sec:framework}). This suggests that the pairwise/listwise formulation itself is a key driver of effectiveness, beyond the quality of the negative sampling strategy. In contrast, the two listwise distillation objectives derived from Rank-DistiLLM \cite{schlattRankDistiLLMClosingEffectiveness2025} (DistillRankNet and ADR-MSE) fall in the middle tier under our unified setup: they are competitive for some backbones and datasets but often trail InfoNCE or MarginMSE. This matters for practitioners: once candidate generation and training protocols are controlled, the benefits of LLM-derived listwise distillation are not universal and should be evaluated against strong non-LLM baselines. Our results further suggest that, under a controlled environment, supervised objectives can be competitive with their distillation-trained counterparts, contradicting recent findings on decoder-only cross-encoders \cite{xu2025distillationversuscontrastivelearning}.

\subsection{Disentangling Objective from Negative Sampling}\label{sec:Negatives_comp}

\begin{figure*}[t]
    \centering
    \vspace{-0.4cm}
    \begin{subfigure}{0.29 \textwidth}
        \centering
        \caption{In Domain (MSM + DL19/20)}
        \includegraphics[width= \textwidth]{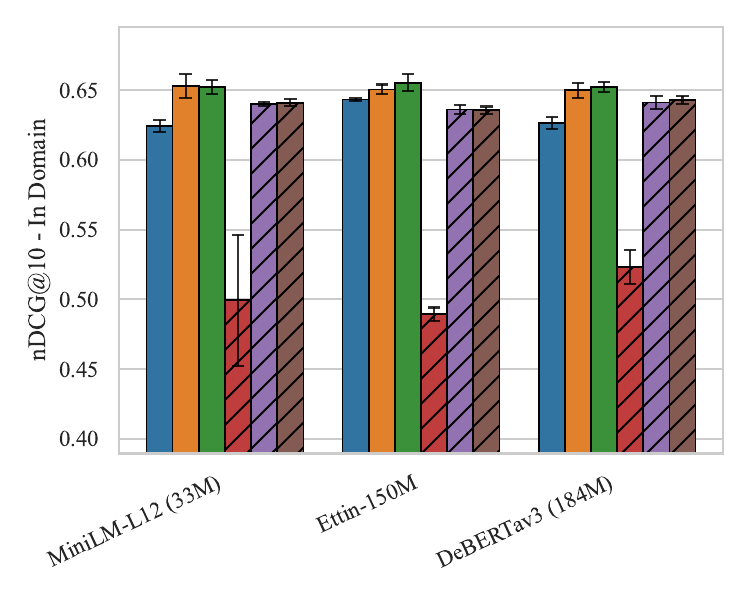}
    \end{subfigure}
    \begin{subfigure}{0.29 \textwidth}
        \centering
        \caption{BEIR-13 (semi OOD)}
        \includegraphics[width= \textwidth]{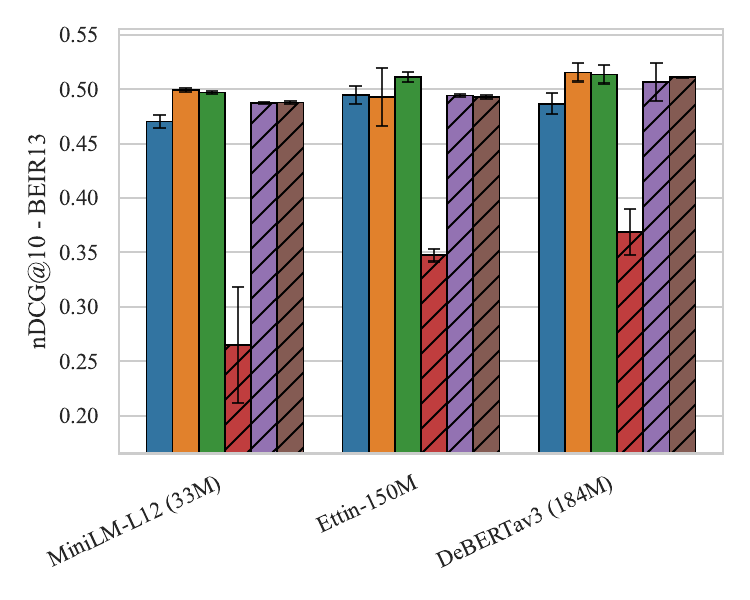}
    \end{subfigure}
    \begin{subfigure}{0.41 \textwidth}
        \centering
        \caption{OOD (LoTTe Search + Robust04)}
        \includegraphics[width= \textwidth]{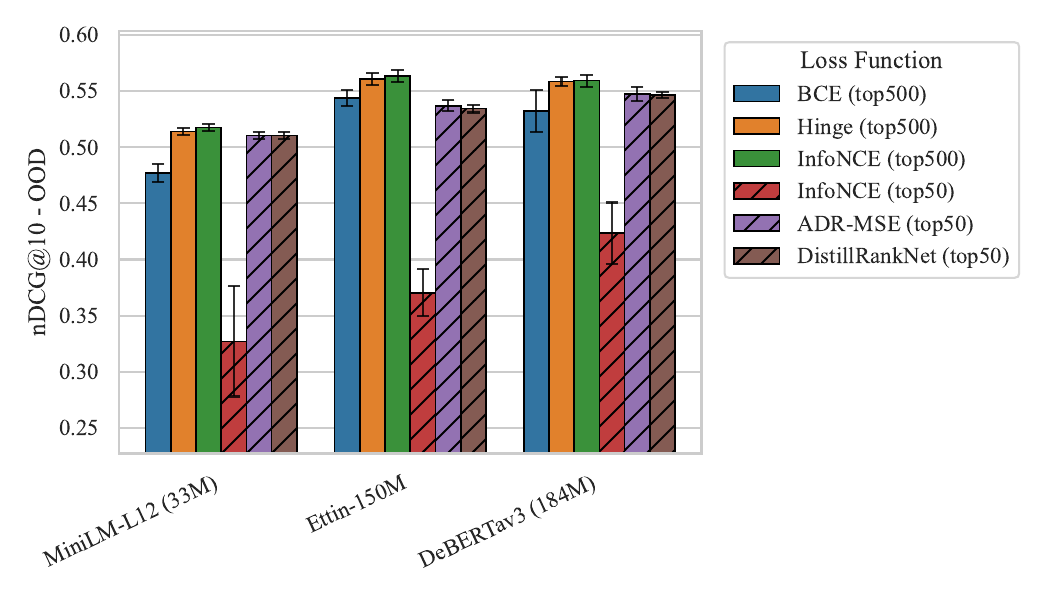}
    \end{subfigure}
    \caption{Comparative evaluation of training objectives with same negatives sampling strategies across three backbones. Solid colors indicate documents sampled in the top 500 from ColBERTv2 \cite{santhanam2022colbertv2} while hatched indicates top 50. We report the mean nDCG@10 across three random seeds for in-domain (ID), BEIR-13 (semi-OOD), and the combined OOD suite (LoTTe Search and Robust04). Error bars indicate the standard deviation across runs, illustrating the relative stability of each training strategy.}
    \label{fig:BackbonesXLossesColbertNeg}
  \vspace{-0.3cm}
\end{figure*}

The training strategies described in \Cref{sec:training_strategies} conflate objectives and sampling strategies for negatives or teacher labels. For instance, BCE and Hinge may not only underperform InfoNCE because of their pairwise formulation---against a stronger listwise formulation---but also because their negatives are sampled from BM25 while InfoNCE benefits from harder negatives mined with ColBERTv2 \cite{schlattRankDistiLLMClosingEffectiveness2025}. Results in \Cref{sec:strategy_impact} evaluate the effect of their combination, but make it difficult to isolate the individual impact of either component. To disentangle these effects, we extend our study with two controlled experiments.

\textbf{Approach.} First, we evaluate pairwise losses (BCE and Hinge) using the same negative sampling as the InfoNCE strategy (\Cref{sec:training_strategies}): following \citet{schlattRankDistiLLMClosingEffectiveness2025}, we sample one positive and 7 negatives from the top-500 documents retrieved by ColBERTv2 \cite{santhanam2022colbertv2} per query, totaling 32 documents per batch with 4 queries.

Second, we compare listwise distillation objectives (ADR-MSE and DistillRankNet) against InfoNCE using the same set of negatives. For each query, \citet{schlattRankDistiLLMClosingEffectiveness2025} re-rank the top-50 documents from ColBERTv2 with their LLM teacher; we add to the positive document 7 negatives sampled randomly from this pool. The positive is still sampled from MS~MARCO qrels, and queries are discarded if no relevant document is in the top-50 pool.

To limit computational costs while ensuring generalizability, we restrict these experiments to three diverse, high-performing backbones: MiniLM, DeBERTaV3, and Ettin-150M. This selection balances effectiveness with diversity in size and architecture.

\textbf{Results.} From \Cref{fig:BackbonesXLossesColbertNeg}, three key trends emerge:

\textit{Backbone consistency.} The relative ordering of objectives holds steady across all three backbones, regardless of size. While DeBERTaV3 provides the highest effectiveness ceiling, the relative gap between training strategies remains essentially constant across backbones, and is most visible on the OOD suite (\Cref{fig:BackbonesXLossesColbertNeg}c).

\textit{Pairwise losses with hard negatives are competitive with LLM distillation.} While InfoNCE still leads when mining negatives from the top-500 documents, standard pairwise losses---once paired with an appropriate negative sampling strategy---are highly competitive. They often match or even outperform more complex listwise distillation methods. For example, on OOD (\Cref{fig:BackbonesXLossesColbertNeg}c), Ettin-150M trained with Hinge loss significantly outperforms its counterparts trained with ADR-MSE or DistillRankNet by three points in mean nDCG@10 (0.56 versus 0.53). This suggests that the signal from a well-distributed set of hard negatives can be more effective than the soft-label supervision provided by an LLM teacher, challenging the necessity of computationally expensive distillation pipelines.

\textit{InfoNCE is sensitive to false negatives in aggressive hard-negative pools.} While InfoNCE (top-500) is a top-tier performer, its effectiveness collapses when restricted to the top-50 sampling pool (red hatched bars), yielding the lowest nDCG@10 scores across all evaluation settings. We attribute this to the higher density of false negatives in the top-50 documents retrieved by ColBERTv2. This echoes the RocketQA study \cite{qu-etal-2021-rocketqa}, which showed the importance of filtering ``negatives'' when mining with a strong model, as MS~MARCO is known to contain many false negatives \cite{thakurHardNegativesHard2025}. The InfoNCE objective appears particularly sensitive to this label noise as the sharp contrastive penalty leads to degraded effectiveness.

\subsection{Backbone Effects and Scaling}\label{sec:backbone_and_scaling}

Applying the same methodology (\Cref{sec:stat_methodology}) across the 18 (objective, scenario) combinations confirms highly significant differences among backbones (Friedman $\chi^2=119.28$, $p<10^{-21}$). Nemenyi ($\alpha=0.05$, CD$=2.83$) reveals a top group comprising ELECTRA (avg.\ rank 1.61), MiniLM-L12 (2.78), Ettin-150M (3.53), DeBERTaV3 (3.72), and BERT-Base (3.89), none of which are significantly different from each other, as confirmed also by the results of TOST ($p<10^{-8}$ across all pairs). At the bottom, Ettin-32M (8.00) and Ettin-17M (9.00) are significantly worse than all base-sized models, except RoBERTa. This ranking matches recent studies, including \citet{schlattRankDistiLLMClosingEffectiveness2025}, who switched to ELECTRA as backbone to fine-tune cross-encoders. At the same time, it suggests that more recent alternatives such as Ettin-150M may now provide an even stronger foundation, while BERT remains surprisingly competitive as a re-ranking model, compared to RoBERTa or the smaller Ettin-68M.

\textit{Objective choice can rival a size upgrade.}
These rankings confirm a pattern visible in \Cref{fig:backbonesXlosses}: the magnitude of objective improvements is often comparable to — and sometimes larger than — the gain from switching to a larger backbone. This is particularly visible in the Ettin scaling suite (17M $\rightarrow$ 32M $\rightarrow$ 68M $\rightarrow$ 150M), where effectiveness increases monotonically with size for nearly all objectives, yet a smaller model trained with a strong objective (e.g., InfoNCE or MarginMSE) can approach or exceed a larger model trained with BCE. Notably, MiniLM-L12 (33M) is statistically indistinguishable from ELECTRA (110M), across training objectives, illustrating that an efficient architecture with good pre-training can match a model three times its size. 
Conversely, scaling a backbone does not compensate for weak supervision choices, especially under OOD evaluation where the BCE gap remains large (from \autoref{tab:results}, Ettin-150M achieves 0.509 nDCG@10 on average on LoTTE when trained with BCE, while the smaller Ettin-32M reaches 0.516 nDCG@10 with MarginMSE).


\subsection{Robustness on OOD}



Across all backbones, MarginMSE and InfoNCE consistently yield the highest effectiveness on LoTTE-Search. InfoNCE on Ettin-150M reaches a peak nDCG@10 of 0.580, the highest value for this benchmark in our study. The effectiveness gap between top-tier objectives remains relatively narrow, suggesting that multiple pairwise/listwise formulations can reliably exploit the structured search intent of this benchmark. Robust04 presents a markedly different picture. With longer documents and older newswire text, the variance across objectives is considerably wider (\Cref{tab:results}): MiniLM-L12 with DistillRankNet achieves the highest score (0.520), and smaller Ettin models benefit disproportionately from distillation. The ranking among top fine-tuning objectives is also less stable, with no single loss consistently dominating across backbones. 

\subsection{Scaling Law for Ettin}

\begin{figure}[t]
    \centering
    \includegraphics[width=0.9\linewidth]{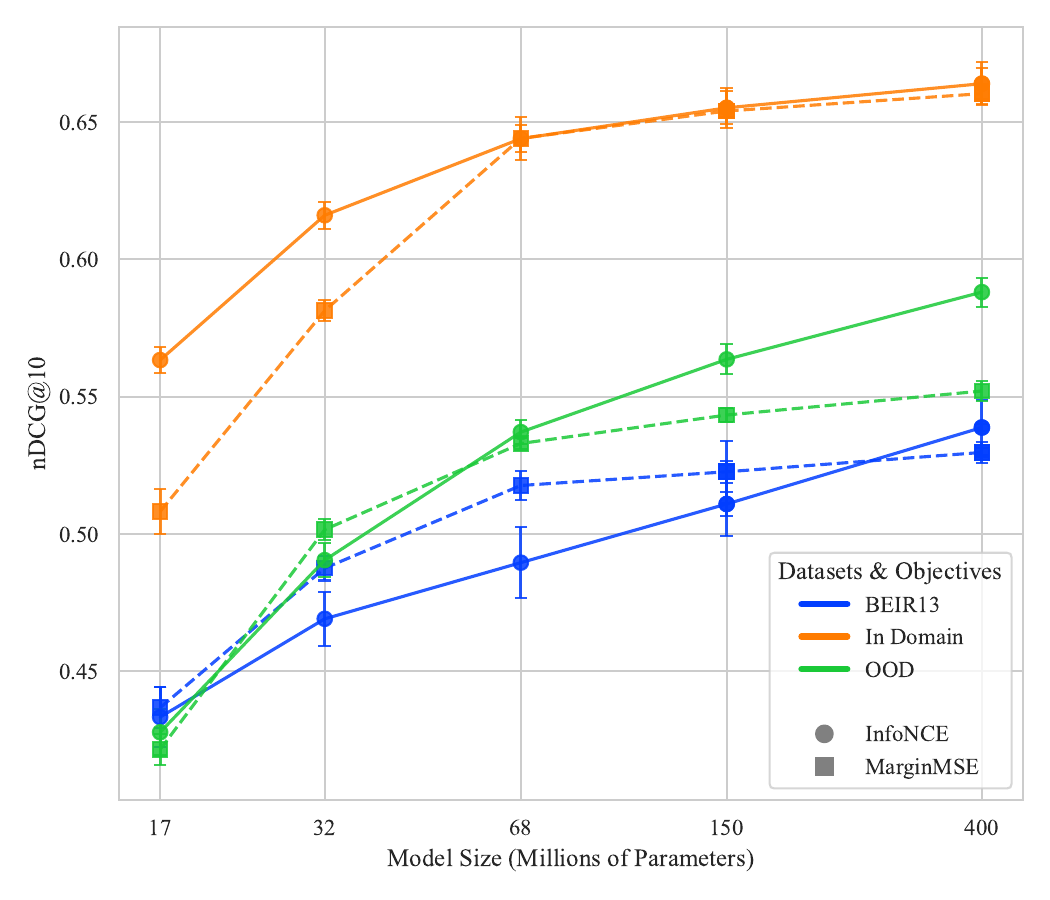}
    \caption{Effectiveness scaling across model sizes for the Ettin \cite{ettin} suite (17M-400M parameters) evaluated ID, on BEIR-13 (Semi OOD), and on OOD. Data series highlight the scaling behavior of two representative distillation objectives: MarginMSE (\citet{hofstatterImprovingEfficientNeural2020}, squares) and InfoNCE (RankDistiLLM  \cite{schlattRankDistiLLMClosingEffectiveness2025}, circles). Error bars represent the standard deviation across random seeds.}
    \label{fig:scaling_ettin}
    \vspace{-.3cm}
\end{figure}

Finally, to study the effect of scaling a model's size across different evaluation settings, we present in \Cref{fig:scaling_ettin} a comparative analysis of model size versus ranking effectiveness based on the Ettin suite. In this visualization, we fix the training strategies to the best ones from our benchmark: MarginMSE (\cite{hofstatterImprovingEfficientNeural2020}, represented by squares) and InfoNCE (using the Rank-DistiLLM \cite{schlattRankDistiLLMClosingEffectiveness2025} strategy, represented by circles), to isolate the effect of scaling the number of parameters, up to the 400M variant, on the re-ranking performance.

The interaction between model capacity and training effectiveness is most clearly visible in the scaling behavior of the Ettin suite, as illustrated in \Cref{fig:scaling_ettin} and detailed in \Cref{tab:results}. We observe a consistent "scaling law" where effectiveness increases monotonically with parameter count, though the rate of improvement varies significantly by objective and evaluation domain. For instance, upgrading from Ettin-17M to Ettin-32M provides a substantial effectiveness gain across all scenarios, such as the increase in BEIR-13 nDCG@10 from 0.433 to 0.469 when using InfoNCE.

However, as models scale up within the suite, we observe diminishing returns: the gain from Ettin-68M to Ettin-150M on OOD is notably narrower, suggesting an effectiveness ``sweet spot'' at mid-range capacities. Crucially, the choice of training objective and backbone can often rival a hardware-intensive size upgrade. As shown in \Cref{tab:results}, a smaller Ettin-32M model trained with MarginMSE (0.426 ID) achieves comparable effectiveness to a larger BERT-Base (110M) model trained with ADR-MSE (0.427 ID), highlighting that supervision strategies can compensate reduced model's size.

Finally, while MarginMSE exhibits superior performance for the smallest models in the suite—specifically at the 17M and 32M parameter scales—a clear crossover point occurs as model capacity increases. For all benchmark categories (ID, BEIR-13, and OOD), InfoNCE emerges as the most effective objective across larger backbones such as Ettin-150M and Ettin-400M. This trend suggests that, as the backbone architecture scales, it gains the capacity to move beyond the simpler signals of pairwise margins and leverage the more complex listwise contrastive signals of InfoNCE, to achieve better generalization.

\section{Conclusion}
We presented a comprehensive, controlled study of training strategies for encoder-only cross-encoders, benchmarking nine architectures across six diverse fine-tuning objectives. By unifying the experimental framework—including candidate generation, hyperparameter optimization, and evaluation protocols—we isolate the true drivers of ranking effectiveness. Our investigation yields three primary takeaways for neural re-ranking:

(i) Objective choice should take precedence over scaling. We demonstrate that the choice of training objective has a statistically significant impact on effectiveness, matching or even exceeding the benefits of architectural scaling. Specifically, InfoNCE or MarginMSE (\Cref{sec:training_strategies}) consistently outperform their counterparts and allow smaller models (e.g., MiniLM, 33M) to match the ranking effectiveness of architectures three times their size.

(ii) Hard negatives matter at least as much as the loss. Disentangling the loss from the negative-sampling pool (\Cref{sec:Negatives_comp}), we find that even a simple pairwise Hinge loss with ColBERTv2 hard negatives matches---and on OOD beats---listwise LLM distillation (DistillRankNet, ADR-MSE \cite{schlattRankDistiLLMClosingEffectiveness2025}). Conversely, InfoNCE collapses when its negatives are drawn from a top-50 pool dense with false negatives, exposing a sensitivity that prior work has overlooked.

(iii) Traditional supervised losses are competitive with LLM distillation. Together, (i) and (ii) imply that the signal from well-distributed hard negatives is more decisive for re-ranker fine-tuning than soft-label supervision from a strong LLM teacher---challenging recent advances in costly LLM-based distillation \cite{schlattRankDistiLLMClosingEffectiveness2025}.

In addition, we release our codebase, the YAML configurations, and the trained model checkpoints along with their training metrics to support replication and extension of these results.

\section{Limitations \& Future Work}

Despite our best efforts to control confounding factors, some limitations remain.

\textbf{Hyperparameters selection.} Due to computational cost, we tuned hyperparameters on proxy models with the BCE loss rather than performing a separate search per backbone and objective. Following \citet{kaplan2020scaling}, we expect this to mostly affect absolute numbers rather than comparative conclusions, though some bias is possible. A full per-backbone, per-objective search would require hundreds of additional experiments; our proxy-based strategy is a practical compromise offering reasonable hyperparameter coverage within a tractable budget.

\textbf{Teacher quality.} Our distillation experiments rely on teacher scores from \citet{schlattRankDistiLLMClosingEffectiveness2025} and \citet{hofstatterImprovingEfficientNeural2020}; the effect of teacher quality remains unexplored.

\textbf{Disentanglement scope.} \Cref{sec:Negatives_comp} disentangles loss formulation from negative-sampling pool for three representative backbones (MiniLM, DeBERTaV3, Ettin-150M); extending this to the full nine-backbone grid would further strengthen the conclusion.

\textbf{OOD evaluation.} Because we use nano-BEIR to select the best checkpoint across seeds, our BEIR results may be biased by its small validation sets (5,000 documents, 50 queries per dataset). While LoTTE and Robust04 remain fully OOD, BEIR comparisons with other work should be read with this caveat in mind.

\textbf{Retrieval setup.} All evaluations use a single first-stage retriever (SPLADE-v3 top-1000), chosen for strong recall; results may differ with other retrievers or candidate depths \citep{Dejean2024ATC}. 

Future work could replicate our experiments with different retrievers, benchmark additional teachers, or scale backbones beyond one billion parameters.

\section*{Acknowledgements}
The authors acknowledge the ANR – FRANCE (French National Research Agency) for its financial support of the GUIDANCE project n°ANR-23-IAS1-0003 as well as the Chaire Multi-Modal/LLM ANR Cluster IA ANR-23-IACL-0007. This work was granted access to the HPC resources of IDRIS under the allocation 2024-AD011015440R1 made by GENCI. The authors also gratefully acknowledge the support of the Centre National de la Recherche Scientifique (CNRS) through a research delegation awarded to J. Mothe.

\section*{GenAI Usage Disclosure}
In this work, generative AI tools were used exclusively for language refinement, to improve the clarity and fluency of the text.

\bibliographystyle{ACM-Reference-Format}
\bibliography{biblio}

\end{document}